\documentstyle[12pt]{article}
\def\beq{\begin{equation}}
\def\eeq{\end{equation}}
\begin{document}
\begin{titlepage}
\begin{center}
  {\Large \bf Theoretical Physics Institute \\ University of Minnesota
    \\} \end{center} \vspace{0.3in}
\begin{flushright}
  TPI-MINN-94/16-T \\ UMN-TH-1252/94 \\ April 1994
\end{flushright}
\vspace{0.4in}
\begin{center}
  {\Large \bf Normalization of QCD corrections in top quark decay\\}
\vspace{0.2in}
{\bf Brian H. Smith \\ }
School of Physics and Astronomy, University of Minnesota, Minneapolis,
    MN 55455 \\
and \\
{\bf M.B. Voloshin \\ }
Theoretical Physics Institute, University of Minnesota, Minneapolis, MN
55455 \\
and\\
Institute for Theoretical and Experimental Physics, Moscow 117259 \\

\vspace{0.2in}
{\bf Abstract \\ }
\end{center}

We discuss the effects of QCD corrections to the on-shell decay $t \to b W$.
We resolve the scale ambiguity using the Brodsky-Lepage-Mackenzie scheme,
and find that the appropriate coupling constant is $\alpha_{s}^{
\overline{MS}} (0.122m_{t})$. The largest long distance contribution comes
from the definition of the on-shell mass of the top quark. We note that QCD
corrections to the electroweak $\rho$ parameter are extremely small when the
$\rho$ parameter is expressed in terms of the top quark width.

\end{titlepage}

There has been much discussion in the literature of QCD corrections to
electroweak processes involving the top quark$^{\cite{dv,sv,loy,sirlin}}$.
One of the motivations for this has been an attempt to extract as much
information as possible from electroweak radiative corrections to
measurements such as $\Gamma(Z \to b\overline{b})$ and the $\rho$ parameter.
With
the measurement of possible signals of top production at the
Tevatron$^{\cite{CDF}}$ and the observation of electroweak radiative
corrections at LEP$^{\cite{norv}}$, there is now the exciting possibility of
using a measured value of $m_{t}$ to extract information about the Higgs boson
or new physics beyond the Standard Mode from precise electroweak data.

In a previous paper$^{\cite{sv}}$, we have discussed QCD effects on
electroweak corrections, specifically the $\rho$ parameter. We showed
that the QCD corrections were dominated by perturbative effects involving
momentum transfers on the order of $m_{t}$. Since any top quarks
appearing in radiative loops have a high virtuality, there are no sizable
$t \overline{t}$ threshold corrections.
The largest nonperturbative effect is associated with expressing the $\rho$
parameter through the top quark mass measured at the pole, $m_{t}$. This is an
uncertainty that arises when a quantity which is well defined at short
distances is written in terms of a long distance parameters and brings about
corrections on the order of $\Lambda_{QCD}/m_{t}$ \footnote{The intrinsic
ambiguity of order $\Lambda_{QCD}$ in the definition of the quark on-shell
mass was also recently emphasized in detail by Bigi {\it et.al.}
$^{\cite{bigi}}$.}.  When the $\rho$ parameter is written in terms of
quantities defined at short distances, all nonperturbative effects are
suppressed by a factor of $(\Lambda_{QCD}/m_{t})^{4}$.

In this paper, we resolve all normalization point ambiguities in the top
decay rate in the order $\alpha_{s}$ by using the method of Brodsky, Lepage,
and Mackenzie$^{\cite{blm}}$. The basic idea of the BLM prescription is that
when one changes the normalization point from $\alpha_{s}(\mu_{1})$ to
$\alpha_{s}(\mu_{2})$, the amplitude will gain a higher order term
proportional to the number of light quark flavors. When a result is
expressed in two different scales, the difference is an $n_{f}$ dependent
higher order in $\alpha_{s}$ term. A poor choice of scale will contain a
large $n_{f}$ dependent higher order corrections. If one wants to find an
appropriate scale to use at order $\alpha_s$, one should calculate the
$n_{f}$ dependent part of the $\alpha_{s}^{2}$ correction. There exists a
scale for which the $n_{f}$ dependent $\alpha_{s}^{2}$ correction is zero.
This scale roughly corresponds to the weighted average of gluon momenta.  It
was argued by BLM that this scale is the physically relevant scale for any
given problem.

In analyzing QCD corrections to the $\rho$ parameter, we found that the
relevant coupling constant according to the BLM prescription was
$\alpha_{s}^{\overline{MS}}(0.154m_{t})$. We will proceed to analyze the QCD
corrections for the on-shell decay rate of $t \to b W$, which is by far
the dominant decay process for the top in the Standard Model.
All calculations are performed in the
limit $m_{t}^{2} >> m_{W}^{2}$. The leading in $\alpha_{s}$ correction
can be found in the literature for finite $m_{W}^{\cite{loy}}$.In
all calculations, the Cabibbo-Kobyashi-Maskawa matrix element $V_{tb}$ is
taken to be unity. In this limit, the width is given to the leading
order in $\alpha_{s}$ by$^{\cite{loy}}$
\begin{equation}
\Gamma_0+\Gamma^{(1)} =
{G_{F}\over{\sqrt{2}}}{m_{t}^{3}\over{8\pi}} \left [
1- {{2 \,\alpha_s}\over{3\pi}} \left ( {{2 \pi^2}\over{3}} - {5\over{2}}
\right )  \right ]~,
\end{equation}
where $\Gamma_0={G_{F}\over{\sqrt{2}}}{m_{t}^{3}\over{8\pi}}$ is the
tree-level width. We wish to resolve the scale ambiguity in this correction
by calculating the $n_{f}$ dependent part of the ${\cal O}(\alpha_{s}^{2})$
correction, and quantify this scale by using the BLM prescription.

The $n_{f}$ dependent virtual correction can be found by considering all one
gluon contributions with an additional vacuum polarization insertion made in
the gluon propagators. There will also be additional bremsstrahlung
contributions arising from the emission of any one of $n_{f}$ pairs of soft
quarks. The calculation of the $n_{f}$ dependent two-loop amplitude is
technically simplified by writing the amplitude as an integral over a
fictitious gluon mass. We will now show how this is done for both the
virtual loop and bremsstrahlung corrections.

To simplify the virtual correction, consider writing the one-loop virtual
contribution, $\Gamma^{(1)}_{virt}$ in terms of an integral over the virtual
gluon momenta,

\beq \Gamma^{(1)}_{virt} = \int{d^{4}k \,\alpha_{s}\, w(k;p_{i})\, D(k^2)},
\label{leadingRate}
\eeq
where $D(k^2)$ is the gluon propagator and $w(k;p_{i})$ is a weight function
depending on the gluon momentum, $k$, and the external momenta, $p_{i}$,
that can be calculated from ordinary Feynman diagrammatic techniques.  The
one-loop gluon vacuum polarization insertion can be made in equation
(\ref{leadingRate}) by replacing the gluon propagator with $D(k^2){\cal
P}_{r}(k^{2})$, where ${\cal P}_{r}(k^{2})$, is the dimensionless gluon
vacuum polarization renormalized at $k^{2}=-m_{t}^{2}$. It is defined in
terms of its unrenormalized counterpart, ${\cal P}(k^{2})$, by a subtraction
at the Euclidean point $k^{2}=-m_{t}^{2}$,

\begin{equation}
{\cal P}_{r}(k^{2}) = {\cal P}(k^{2}) - {\cal P}(-m_{t}^{2})
\label{firstVirt}
\end{equation}

Renormalizing in this way corresponds to the Brodsky - Lepage - Mackenzie
so-called $V$ scheme (in which $\alpha_s(Q)$ is normalized by the Coulomb
potential between two infinitely massive color objects at momentum transfer
$Q$).  The $n_{f}$ dependent part of the two-loop virtual correction is,
when expressed in the same form as equation ($ \ref{firstVirt}$),

\begin{equation}
 \delta\Gamma^{(2)}_{virt} = \int{d^{4}k \, \alpha_{s}(m_t) \,
 w(k; p_i)\,
  {{\cal P}_{r}(k^2)\over{k^2}}}.
\label{nfterm}
\end{equation}

The vacuum polarization of the gluon can be written as an integral over its
imaginary part through the dispersion relation with a subtraction at $k^{2}
= -m_{t}^{2}$,
\begin{equation}
  {\cal P}_{r}(k^2) = {1\over{\pi}} \int_{0}^{\infty} {\rm Im} {\cal P}
(\mu^{2}) \left [ {1\over{\mu^{2}-k^{2}}} - {1\over{\mu^{2}+m_{t}^{2}}}
             \right ] d\mu^{2}.
\label{disperseSum}
\end{equation}

By adding and subtracting $1/\mu^{2}$ in the weight function,
the dispersion relation in equation ($\ref{disperseSum}$) becomes,
\begin{equation}
{\cal P}_{r}(k^{2}) = -{k^{2}\over{\pi}} \int_{0}^{\infty} {{\rm Im} {\cal
    P}(\mu^{2})\over{\mu^{2}(k^{2}-\mu^{2})}} d\mu^{2} +
 {m_t^{2}\over{\pi}} \int_{0}^{\infty} {{\rm Im} {\cal
    P}(\mu^{2})\over{\mu^{2}(\mu^{2}+m_{t}^{2})}} d\mu^{2}.
\label{disperse}
\end{equation}

The first term in the dispersion integral
(\ref{disperse}) when taken together with equation (\ref{nfterm}) has the form
of an integral over the gluon mass. The denominator in the first
integrand in equation
($\ref{disperse}$) is a massive gluon propagator. The $n_{f}$ dependent
part of the virtual two-loop contribution to the width can be rewritten as
an integral over the gluon mass in the one-loop contribution,
\begin{equation}
\delta\Gamma^{(2)}_{virt} = -{1\over{\pi}} \int_{0}^{\infty} \left [
\Gamma^{(1)}_{virt}(\mu) - {m_{t}^{2}\over{\mu^{2}+m_{t}^{2}}}
\Gamma^{(1)}_{virt}(0) \right ]
{{{\rm Im} {\cal P}}(\mu^2)\over {\mu^2}} d\mu^{2},
\label{virtWidth}
\end{equation}
where $\Gamma^{(1)}_{virt}(\mu)$ is the virtual contribution of the one-loop
width calculated with a gluon of mass $\mu$. Although
$\Gamma^{(1)}_{virt}(\mu)$ is divergent as $\mu^{2} \to 0$, the second term in
the integrand of equation ($\ref{virtWidth}$) should be understood as having a
small regularizing gluon mass. This infrared divergence will be cancelled by
the bremsstrahlung terms, below.

We now show that an analogous relation is true for the
bremsstrahlung contribution. The $n_{f}$ dependent part of the ${\cal
O}(\alpha_{s})$ bremsstrahlung correction receives contributions from two
sources. The first corresponds to the emission of two soft tag quarks.
The second comes from making a vacuum polarization insertion in the final
state gluon propagator of normal one gluon bremsstrahlung, and
corresponds to the renormalization of the gluon wave function in the process
$t \to b W g$. The former type of process, quark bremsstrahlung, shall be
discussed first.

The four body phase space can be
written as the integral over the product of a three body phase space
and a two body phase space, so that the first bremsstrahlung contribution to
the width is given by,

\begin{equation}
  \Gamma^{(2)}_{q-brem} = \int_{0}^{m_{t}^2} {{\cal M}^{2}\over{2 m_{t}}}
 \, d\tau_{2}(\mu) \, d\tau_{3}(m_{t})\, {d\mu^{2}\over{2\pi}},
\label{twoPhase}
\end{equation}
where ${\cal M}^{2}$ is the square of the matrix element,
$d\tau_{2}(\mu)$ is the two body phase space of the two tag quarks
with center of mass energy $\mu$, and $d\tau_{3}(m_{t})$ is the three
body phase space of the $W$ boson, the b quark, and an on-shell
``gluon'' of mass $\mu$.

The phase space integral can be simplified  from the form of Equation
($\ref{twoPhase}$) . Integration over the three body phase space will result in
a factor of $\Gamma^{(1)}_{brem}(\mu)$, the width of the process
$t \to W\, b\, g(\mu)$, where $g(\mu)$ is a gluon of mass $\mu$. The
integral over the two body phase space of the light quarks can be
written in terms of the imaginary part of the gluon polarization
operator. The quark bremsstrahlung contribution is,
\begin{equation}
  \Gamma^{(2)}_{q-brem} = -{1\over{\pi}}\int_{0}^{m_{t}^{2})}
  {\Gamma^{(1)}_{brem}(\mu){{\rm Im} P}(\mu^2) d\mu^{2}}~.
\label{qbremWidth}
\end{equation}
(It should be noted in this connection that in the standard definition of
the vacuum polarization  ${\cal P}(k^2)$ as entering the exact propagator as
$D(k^2)=\left ( k^2 \, (1- {\cal P}(k^2)) \right )^{-1}$, the
contribution of a quark pair to the imaginary part ${\rm Im}{\cal P}$ is
negative.)

The final contribution comes from the gluon wave function
renormalization in the single gluon
bremsstrahlung.  The $n_{f}$ dependent piece is calculated by making a
vacuum polarization insertion in the final gluon propagator appearing in the
soft bremsstrahlung process $t \to b W g$. The correction to the width is
$\Gamma^{(1)}_{brem}(0) {\cal P}(0)$ and can be written using the dispersion
integral ($\ref{disperse}$) evaluated at $k^{2} = 0$.
\begin{equation}
\delta\Gamma^{(2)}_{g-brem} = {1\over{\pi}} \int_{0}^{\infty}
  \Gamma^{(1)}_{brem} {\rm Im} {\cal
P}(\mu^{2}){m_{t}^{2}\over{\mu^{2}(\mu^{2}+m_{t}^{2})}} d\mu^{2},
\label{gbremWidth}
\end{equation}
where $\Gamma^{(1)}_{brem}(0)$ is the tree-level
single gluon bremsstrahlung partial width.

Like the virtual contribution, equation ($\ref{virtWidth}$), the
quark bremsstrahlung contribution, equation ($\ref{qbremWidth}$), has the
form of an integral over the gluon mass. Note that the upper limit of the
integration in equation ($\ref{qbremWidth}$) is $m_{t}^{2}$ instead of
infinity. However, if the decay rate, $\Gamma^{(1)}_{brem}(\mu^{2})$,
is understood as containing a step function, i.e.
$\Gamma^{(1)}_{brem}(\mu^{2}) = 0$ for all $\mu^{2} > m_{t}^{2}$, the
total $n_{f}$ dependent two-loop contribution becomes a single integral over
the gluon mass,
 \begin{equation}
  \delta\Gamma^{(2)} = -{1\over{\pi}}\int_{0}^{m_{t}^2}{ \left (
    \Gamma^{(1)}_{brem}(\mu) + \Gamma^{(1)}_{virt}(\mu)
    -  {m_{t}^{2}\over{(\mu^{2}+m_{t}^{2})}} \Gamma^{(1)}(0) \right ){{\rm Im}
      P}(\mu^2) {d\mu^{2}\over{\mu^{2}}}},
\label{massInt}
\end{equation}
While individual contributions may be infrared divergent, the integral in
equation ($\ref{massInt}$) is well behaved both in the infrared and
ultraviolet regions.

The procedure for calculating the $n_{f}$ dependent correction is now
clear. First, we calculate the leading virtual and
bremsstrahlung corrections with a fictitious gluon mass, $\mu$. This
gluon mass will also serve the purpose of regularizing any
infrared divergences that might otherwise arise in intermediate
results. Next, we perform a weighted integration over the gluon mass.
The weight function is given in equation ($\ref{massInt}$).

The virtual massive gluon correction to the width,
$\Gamma^{(1)}_{virt}(\mu^{2})$, is found by calculating diagrams
corresponding to the dressing of the vertex, the initial state
top quark propagator, and the final state b quark propagator with
a massive gluon. The explicit result of this calculation is,
\begin{eqnarray}
  \Gamma^{(1)}_{virt}(\mu) = \Gamma_{0} {2\over{3}}
{\alpha_{s}\over{\pi}} [ -{5\over{4}}(3+2x) + {x(28 + 10x
    - 5x^{2})\over{2 \sqrt{x(4-x)}}}\, \arctan{\sqrt{4-x\over{x}}} -
  \nonumber \\ {5 \over{4}} (2-x^2)\, \log{x} - 2 (1+x)\log{\left
[{1\over{2}}(x+\sqrt{x(x-4)})\right ]} \log{\left [
  {1\over{2}}(x-\sqrt{x(x-4)}) \right ] } ]~,
\label{fvirt}
\end{eqnarray}
where $x$ is the dimensionless gluon to top mass ratio,
${\mu^{2}/{m_{t}^{2}}}$, and $\Gamma_{0}$ is the tree level top width.

The bremsstrahlung term is calculated by considering
the emission of a massive gluon from either of the quarks in t decay. The
partial width of this process is,
\begin{eqnarray}
\Gamma^{(1)}_{brem}(\mu) = \Gamma_{0} {2\over{3}} {\alpha_{s}\over{\pi}}
[ {1\over{4}}(25 - 18x -7 x^2) - 6(1+x)\left ( \arctan {\left [
{-\sqrt{3}+\sqrt{x(4-x)}\over{x-3}} \right ]}\right )^{2} + \nonumber \\
{x(20+2x-x^2)\over{2\sqrt{x(4-x)}}} \arctan {\left [ {(1-x)
    \sqrt{x(4-x)}\over{x(x-3)}} \right ]} + {1\over{4}}(10+x^2)\log{x}
+{1\over{2}}(1+x)(\log{x})^{2}~.
\label{fbrem}
\end{eqnarray}

The sum of the expressions ($\ref{fvirt}$) and ($\ref{fbrem}$) is finite
both in the infrared and ultraviolet regions. It is a simple matter to carry
out the integration numerically. For a single light tag fermion in an
Abelian theory, the imaginary part of the gluon vacuum polarization, ${\cal
P}(\mu^{2})$ is $-\alpha/3$.  The proper coefficient for QCD can be obtained
by replacing ${\rm Im} {\cal P}(k^{2})$ with $\alpha_{s} b/4$, where $b=11-{2
\over 3} n_{f}$ is the first coefficient in the QCD beta function.  The
integral in equation ($\ref{massInt}$) can be evaluated numerically, which
gives the decay rate as

\begin{equation}
\Gamma_{t} = \Gamma_{0} \left [ 1 -
{2\over{3\pi}}\alpha^{V}_{s}(m_{t}) \left ( 2
{\pi^{2}\over{3}} - {5\over{2}} \right ) \left ( 1 + {{b
\alpha_s}\over{4\pi}} 2.54 \right ) \right ]~.
\label{potAnswer}
\end{equation}

Within the BLM prescription, the appropriate normalization point for
$\alpha_{s}$ is chosen so that there are no $n_{f}$ dependent terms of
${\cal O}(\alpha_{s}^{2})$. The normalization point can be found by
rescaling $\alpha_{s}$ equation ($\ref{potAnswer}$)  by,
\begin{equation}
\alpha^{V}_{s}(m_{t}) = \alpha^{V}_{s}(Q) \left [ 1 + {b\over{4\pi}}
\alpha_{s}(Q) \log{Q^{2}\over{m_{t}^{2}}} \right ],
\end{equation}
and choosing a scale $Q$ such that
the $n_{f}$ dependent term in equation ($\ref{potAnswer}$) is cancelled by the
$n_{f}$ dependent term coming from the rescaling of $\alpha_{s}$.
Numerically, this scale is  $0.281 m_{t}$. Equation ($\ref{potAnswer}$) can
be written with this new scale,
\begin{equation}
\Gamma_{t} = \Gamma_{0} \left [ 1- {2\over{3\pi}} \left ( {{2
\pi^{2}}\over{3}} - {5\over{2}} \right ) \alpha^{V}_{s}(0.281 m_{t}) \right
]
\label{newScale}
\end{equation}

The coupling constant in the above equation is written in terms of
$\alpha^{V}_{s}$, which is derived from the potential between two
infinitely massive quarks.
This coupling constant can be related to the more conventional
$\overline{MS}$ scheme by $\alpha^{V}_{s}(k) =
\alpha_{s}^{\overline{MS}}(e^{-5/6}k)~^{\cite{blm}}$. The coupling constant
in equation ($\ref{newScale}$) is then equal to
$\alpha^{\overline{MS}}_{s}(0.122 m_{t})$.

It is interesting to note that this scale is close to the scale of
$0.154 m_{t}$ which was
shown in a previous work$^{\cite{sv}}$ to characterize QCD correction to
the electroweak $\rho$ parameter. We recall the result that the $\rho$
parameter  is, with a properly normalized one gluon correction,
\begin{equation}
\delta \rho \equiv {{3\, G_{F} m_t^2}\,\over{8\pi^{2} \sqrt{2}}} \left [ 1 -
{2\over{3\pi}} ({\pi^{2}\over{3}}+1) \alpha_{s}^{\overline{MS}}(0.154 m_{t})
\right ],
\label{deltaRho}
\end{equation}
to leading order in $m_{t}^{2}/m_{w}^{2}$.

QCD corrections to the top width and the $\rho$ parameter are physically
characterized by momenta transfer scales on the order of $m_{t}$.
When these physical quantities are written in terms of a mass defined
at long distances (like the pole mass used above), corrections are
introduced corresponding to gluon effects on momenta scales smaller than
$m_{t}$. This has the effect of bringing the normalization point of
the coupling constant down to a lower scale. As discussed by
Sirlin$^{\cite{sirlin}}$, the QCD correction in equation ($\ref{deltaRho}$)
can be rewritten in terms of the running $\overline{MS}$ mass,
$m_{t}^{\overline{MS}}$,
normalized at the scale $m_{t}$. The corrections to the $\rho$ parameter,
when expressed through this running $\overline{MS}$ mass, contain only
quantities defined at short distances, and any long distance nonperturbative
effects are suppressed by a factor of ${\cal
O}((\Lambda_{QCD}/m_t)^{4})$.

While this form may be theoretically more palatable, the quantity
$m_{t}^{\overline{MS}}$ is not easily interpreted in a physical way. The
numerical closeness of the QCD corrections to $\Gamma_{t}$ and $\Delta\rho$
suggests that it may be prudent to express $\Delta\rho$ in terms of the
top quark decay rate. Equations ($\ref{newScale}$) and ($\ref{deltaRho}$) can
be combined in the relation,
\begin{equation}
\Delta\rho = {\Gamma_{t}\over{m_{t}}} {3\over{\pi}}  \left [ 1 -
  {2\over{3\pi}} \left ( {7\over{2}}-{\pi^{2}\over{3}}  \right )
    \alpha_{s}^{\overline{MS}}(0.91 m_{t})  \right ] .
\end{equation}

The order $\alpha_{s}$ correction is numerically very small (the relative
correction is $0.0446 \alpha_{s}$). Writing the $\rho$ parameter in this
way is convenient because the leading QCD corrections are tiny, the
normalization
point of $\alpha_{s}$ is appropriate for momenta transfers on the
order of $m_{t}$.

Both the $\rho$ parameter and the ratio $\Gamma_t/m_t$ are physical
quantities which only contain corrections that enter through short distance
physics. If these short distance corrections are written in terms of
quantities defined at longer distances like the pole mass, there is some
uncertainty of ${\cal O}(\Lambda_{QCD}/m_{t})$ associated with relating a
long range parameter to a short range one. This uncertainty in minimized by
writing short distance parameters in terms of each other, like writing the
$\rho$ parameter in terms of $\Gamma_{t}/m_t$.

This work was supported in part by DOE grant DOE-AC-02-83ER40105. BHS is
supported by the University of Minnesota Doctoral Dissertation Fellowship
program.

\end{document}